# Efficient Processing of k Nearest Neighbor Joins using MapReduce

Wei Lu        Yanyan Shen        Su Chen        Beng Chin Ooi

National University of Singapore

{luwei1,shenyanyan,chensu,ooibc}@comp.nus.edu.sg

## ABSTRACT

$k$ nearest neighbor join ($k$NN join), designed to find $k$ nearest neighbors from a dataset $S$ for every object in another dataset $R$, is a primitive operation widely adopted by many data mining applications. As a combination of the $k$ nearest neighbor query and the join operation, $k$NN join is an expensive operation. Given the increasing volume of data, it is difficult to perform a $k$NN join on a centralized machine efficiently. In this paper, we investigate how to perform $k$NN join using MapReduce which is a well-accepted framework for data-intensive applications over clusters of computers. In brief, the mappers cluster objects into groups; the reducers perform the $k$NN join on each group of objects separately. We design an effective mapping mechanism that exploits pruning rules for distance filtering, and hence reduces both the shuffling and computational costs. To reduce the shuffling cost, we propose two approximate algorithms to minimize the number of replicas. Extensive experiments on our in-house cluster demonstrate that our proposed methods are efficient, robust and scalable.

## 1. INTRODUCTION

$k$ nearest neighbor join ($k$NN join) is a special type of join that combines each object in a dataset $R$ with the $k$ objects in another dataset $S$ that are closest to it. $k$NN join typically serves as a primitive operation and is widely used in many data mining and analytic applications, such as the $k$-means and $k$-medoids clustering and outlier detection [5, 12].

As a combination of the $k$ nearest neighbor ($k$NN) query and the join operation, $k$NN join is an expensive operation. The naive implementation of $k$NN join requires scanning $S$ once for each object in $R$ (computing the distance between each pair of objects from $R$ and $S$), easily leading to a complexity of $O(|R| \cdot |S|)$. Therefore, considerable research efforts have been made to improve the efficiency of the $k$NN join [4, 17, 19, 18]. Most of the existing work devotes themselves to the design of elegant indexing techniques for avoiding scanning the whole dataset repeatedly and for pruning as many distance computations as possible.

All the existing work [4, 17, 19, 18] is proposed based on the centralized paradigm where the $k$NN join is performed on a sin-gle, centralized server. However, given the limited computational capability and storage of a single machine, the system will eventually suffer from performance deterioration as the size of the dataset increases, especially for multi-dimensional datasets. The cost of computing the distance between objects increases with the number of dimensions; and the curse of the dimensionality leads to a decline in the pruning power of the indexes.

Regarding the limitation of a single machine, a natural solution is to consider parallelism in a distributed computational environment. MapReduce [6] is a programming framework for processing large scale datasets by exploiting the parallelism among a cluster of computing nodes. Soon after its birth, MapReduce gains popularity for its simplicity, flexibility, fault tolerance and scalability. MapReduce is now well studied [10] and widely used in both commercial and scientific applications. Therefore, MapReduce becomes an ideal framework of processing $k$NN join operations over massive, multi-dimensional datasets.

However, existing techniques of $k$NN join cannot be applied or extended to be incorporated into MapReduce easily. Most of the existing work rely on some centralized indexing structure such as the B$^+$-tree [19] and the R-tree [4], which cannot be accommodated in such a distributed and parallel environment directly.

In this paper, we investigate the problem of implementing $k$NN join operator in MapReduce. The basic idea is similar to the hash join algorithm. Specifically, the mapper assigns a key to each object from $R$ and $S$; the objects with the same key are distributed to the same reducer in the shuffling process; the reducer performs the $k$NN join over the objects that have been shuffled to it. To guarantee the correctness of the join result, one basic requirement of data partitioning is that for each object $r$ in $R$, the $k$ nearest neighbors of $r$ in $S$ should be sent to the same reducer as $r$ does, i.e., the $k$ nearest neighbors should be assigned with the same key as $r$. As a result, objects in $S$ may be replicated and distributed to multiple reducers. The existence of replicas leads to a high shuffling cost and also increases the computational cost of the join operation within a reducer. Hence, a good mapping function that minimizes the number of replicas is one of the most critical factors that affect the performance of the $k$NN join in MapReduce.

In particular, we summarize the contributions of the paper as follows.

- We present an implementation of $k$NN join operator using MapReduce, especially for large volume of multi-dimensional data. The implementation defines the mapper and reducer jobs and requires no modifications to the MapReduce framework.

- We design an efficient mapping method that divides objects into groups, each of which is processed by a reducer to

perform the $k$NN join. First, the objects are divided into partitions based on a Voronoi diagram with carefully selected pivots. Then, data partitions (i.e., Voronoi cells) are clustered into groups only if the distances between them are restricted by a specific bound. We derive a distance bound that leads to groups of objects that are more closely involved in the $k$NN join.

- We derive a cost model for computing the number of replicas generated in the shuffling process. Based on the cost model, we propose two grouping strategies that can reduce the number of replicas greedily.

- We conduct extensive experiments to study the effect of various parameters using two real datasets and some synthetic datasets. The results show that our proposed methods are efficient, robust, and scalable.

The remainder of the paper is organized as follows. Section 2 describes some background knowledge. Section 3 gives an overview of processing $k$NN join in MapReduce framework, followed by the details in Section 4. Section 5 presents the cost model and grouping strategies for reducing the shuffling cost. Section 6 reports the experimental results. Section 7 discusses related work and Section 8 concludes the paper.

## 2. PRELIMINARIES

In this section, we first define $k$NN join formally and then give a brief review of the MapReduce framework. Table 1 lists the symbols and their meanings used throughout this paper.

### 2.1 $k$NN Join

We consider data objects in an $n$-dimensional metric space $\mathcal{D}$. Given two data objects $r$ and $s$, $|r, s|$ represents the distance between $r$ and $s$ in $\mathcal{D}$. For the ease of exposition, the Euclidean distance ($L_2$) is used as the distance measure in this paper, i.e.,

$$|r, s| = \sqrt{\sum_{1 \le i \le n} (r[i] - s[i])^2}, \qquad (1)$$

where $r[i]$ (resp. $s[i]$) denotes the value of $r$ (resp. $s$) along the $i^{th}$ dimension in $\mathcal{D}$. Without loss of generality, our methods can be easily applied to other distance measures such as the Manhattan distance ($L_1$), and the maximum distance ($L_\infty$).

DEFINITION 1. ($k$ **nearest neighbors**) *Given an object $r$, a dataset $S$ and an integer $k$, the $k$ nearest neighbors of $r$ from $S$, denoted as $KNN(r, S)$, is a set of $k$ objects from $S$ that $\forall o \in KNN(r, S), \forall s \in S - KNN(r, S), |o, r| \le |s, r|$.*

DEFINITION 2. ($k$**NN join**) *Given two datasets $R$ and $S$ and an integer $k$, $k$NN join of $R$ and $S$ (denoted as $R \ltimes_{KNN} S$, abbreviated as $R \ltimes S$), combines each object $r \in R$ with its $k$ nearest neighbors from $S$. Formally,*

$$R \ltimes S = \{(r, s) | \forall r \in R, \forall s \in KNN(r, S)\} \qquad (2)$$

According to Definition 2, $R \ltimes S$ is a subset of $R \times S$. Note that $k$NN join operation is asymmetric, i.e., $R \ltimes S \ne S \ltimes R$. Given $k \le |S|$, the cardinality of $|R \ltimes S|$ is $k \times |R|$. In the rest of this paper, we assume that $k \le |S|$. Otherwise, $k$NN join degrades to the cross join and just generates the result of Cartesian product $R \times S$.

**Table 1: Symbols and their meanings**

| Symbol | Definition |
| --- | --- |
| $\mathcal{D}$ | an $n$-dimensional metric space |
| $R$ (resp. $S$) | an object set $R$ (resp. $S$) in $\mathcal{D}$ |
| $r$ (resp. $s$) | an object, $r \in R$ (resp. $s \in S$) |
| $|r, s|$ | the distance from $r$ to $s$ |
| $k$ | the number of near neighbors |
| $KNN(r, S)$ | the $k$ nearest neighbors of $r$ from $S$ |
| $R \ltimes S$ | $k$NN join of $R$ and $S$ |
| $\mathbb{P}$ | a set of pivots |
| $p_i$ | a pivot in $\mathbb{P}$ |
| $p_r$ | the pivot in $\mathbb{P}$ that is closest to $r$ |
| $P_i^R$ | the partition of $R$ that corresponds to $p_i$ |
| $p_i.d_j$ | the $j^{\text{th}}$ smallest distance of objects in $P_i^S$ to $p_i$ |
| $U(P_i^R)$ | $\max\{|r, p| | \forall r \in P_i^R\}$ |
| $L(P_i^R)$ | $\min\{|r, p| | \forall r \in P_i^R\}$ |
| $T_R$ | the summary table for partitions in $R$ |
| $N$ | the number of reducers |

### 2.2 MapReduce Framework

MapReduce [6] is a popular programming framework to support data-intensive applications using shared-nothing clusters. In MapReduce, input data are represented as key-value pairs. Several functional programming primitives including Map and Reduce are introduced to process the data. Map function takes an input key-value pair and produces a set of intermediate key-value pairs. MapReduce runtime system then groups and sorts all the intermediate values associated with the same intermediate key, and sends them to the Reduce function. Reduce function accepts an intermediate key and its corresponding values, applies the processing logic, and produces the final result which is typically a list of values.

Hadoop is an open source software that implements the MapReduce framework. Data in Hadoop are stored in HDFS by default. HDFS consists of multiple DataNodes for storing data and a master node called NameNode for monitoring DataNodes and maintaining all the meta-data. In HDFS, imported data will be split into equal-size chunks, and the NameNode allocates the data chunks to different DataNodes. The MapReduce runtime system establishes two processes, namely JobTracker and TaskTracker. The JobTracker splits a submitted job into map and reduce tasks and schedules the tasks among all the available TaskTrackers. TaskTrackers will accept and process the assigned map/reduce tasks. For a map task, the TaskTracker takes a data chunk specified by the JobTracker and applies the `map()` function. When all the `map()` functions complete, the runtime system groups all the intermediate results and launches a number of reduce tasks to run the `reduce()` function and produce the final results. Both `map()` and `reduce()` functions are specified by the user.

### 2.3 Voronoi Diagram-based Partitioning

Given a dataset $O$, the main idea of Voronoi diagram-based partitioning is to select $M$ objects (which may not belong to $O$) as pivots, and then split objects of $O$ into $M$ disjoint partitions where each object is assigned to the partition with its closest pivot [1]. In this way, the whole data space is split into $M$ "generalized Voronoi cells". Figure 1 shows an example of splitting objects into 5 partitions by employing the Voronoi diagram-based partitioning. For

---

[1]In particular, if there exist multiple pivots that are closest to an object, then the object is assigned to the partition with the smallest number of objects.

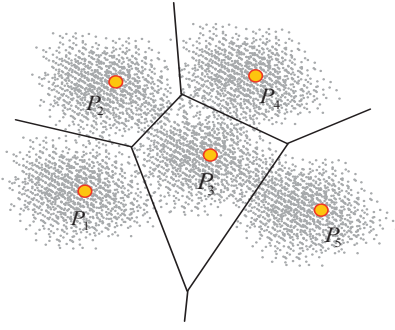

**Figure 1: An example of data partitioning**

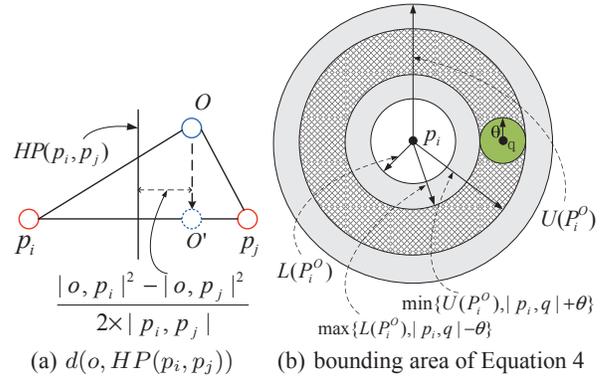

(a) $d(o, HP(p_i, p_j))$      (b) bounding area of Equation 4

**Figure 2: Properties of data partitioning**

the sake of brevity, let $\mathbb{P}$ be the set of pivots selected. $\forall p_i \in \mathbb{P}$, $P_i^O$ denotes the set of objects from $O$ that takes $p_i$ as their closest pivot. For an object $o$, let $p_o$ and $P_o^O$ be its closest pivot and the corresponding partition respectively. In addition, we use $U(P_i^O)$ and $L(P_i^O)$ to denote the maximum and minimum distance from pivot $p_i$ to the objects of $P_i^O$, i.e., $U(P_i^O) = \max\{|o, p_i| | \forall o \in P_i^O\}$, $L(P_i^O) = \min\{|o, p_i| | \forall o \in P_i^O\}$.

**DEFINITION 3. (Range Selection)** *Given a dataset $O$, an object $q$, and a distance threshold $\theta$, range selection of $q$ from $O$ is to find all objects (denoted as $\tilde{O}$) of $O$, such that $\forall o \in \tilde{O}, |q, o| \leq \theta$.*

By splitting the dataset into a set of partitions, we can answer range selection queries based on the following theorem.

**THEOREM 1. [8]** *Given two pivots $p_i$, $p_j$, let $HP(p_i, p_j)$ be the generalized hyperplane, where any object $o$ lying on $HP(p_i, p_j)$ has the equal distance to $p_i$ and $p_j$. $\forall o \in P_j^O$, the distance of $o$ to $HP(p_i, p_j)$, denoted as $d(o, HP(p_i, p_j))$ is:*

$$d(o, HP(p_i, p_j)) = \frac{|o, p_i|^2 - |o, p_j|^2}{2 \times |p_i, p_j|} \quad (3)$$

Figure 2(a) shows distance $d(o, HP(p_i, p_j))$. Given object $q$, its belonging partition $P_q^O$, and another partition $P_i^O$, according to Theorem 1, it is able to compute the distance from $q$ to $HP(p_q, p_i)$. Hence, we can derive the following corollary.

**COROLLARY 1.** *Given a partition $P_i^O$ and $P_i^O \neq P_q^O$, if we can derive $d(q, HP(p_q, p_i)) > \theta$, then $\forall o \in P_i^O, |q, o| > \theta$.*

Given a partition $P_i^O$, if we get $d(q, HP(p_q, p_i)) > \theta$, according to Corollary 1, we can discard all objects of $P_i^O$. Otherwise, we check partial objects of $P_i^O$ based on Theorem 2.

**THEOREM 2. [9, 20]** *Given a partition $P_i^O$, $\forall o \in P_i^O$, the necessary condition that $|q, o| \leq \theta$ is:*

$$\max\{L(P_i^O), |p_i, q| - \theta\} \leq |p_i, o| \leq \min\{U(P_i^O), |p_i, q| + \theta\} \quad (4)$$

Figure 2(b) shows an example of the bounding area of Equation 4. To answer range selections, we only need to check objects that lie in the bounding area of each partition.

## 3. AN OVERVIEW OF KNN JOIN USING MAPREDUCE

In MapReduce, the mappers produce key-value pairs based on the input data; each reducer performs a specific task on a group of pairs with the same key. In essence, the mappers do something similar to (typically more than) the hashing function. A naive and straightforward idea of performing $k$NN join in MapReduce is similar to the hash join algorithm.

Specifically, the map() function assigns each object $r \in R$ a key; based on the key, $R$ is split into disjoint subsets, i.e., $R = \bigcup_{1 \leq i \leq N} R_i$, where $R_i \bigcap R_j = \emptyset, i \neq j$; each subset $R_i$ is distributed to a reducer. Without any pruning rule, the entire set $S$ has to be sent to each reducer to be joined with $R_i$; finally $R \ltimes S = \bigcup_{1 \leq i \leq N} R_i \ltimes S$.

In this scenario, there are two major considerations that affect the performance of the entire join process.

1. The shuffling cost of sending intermediate results from mappers to reducers.

2. The cost of performing the $k$NN join on the reducers.

Obviously, the basic strategy is too expensive. Each reducer performs $k$NN join between a subset of $R$ and the entire $S$. Given a large population of $S$, it may go beyond the capability of the reducer. An alternative framework [21], called H-BRJ, splits both $R$ and $S$ into $\sqrt{N}$ disjoint subsets, i.e., $R = \bigcup_{1 \leq i \leq \sqrt{N}} R_i$, $S = \bigcup_{1 \leq j \leq \sqrt{N}} S_j$. Similarly, the partitioning of $R$ and $S$ in H-BRJ is performed by the map() function; a reducer performs the $k$NN join between a pair of subsets $R_i$ and $S_j$; finally, the join results of all pairs of subsets are merged and $R \ltimes S = \bigcup_{1 \leq i, j \leq \sqrt{N}} R_i \ltimes S_j$. In H-BRJ, $R$ and $S$ are partitioned into equal-sized subsets on a random basis.

While the basic strategy can produce the join result using one MapReduce job, H-BRJ requires two MapReduce jobs. Since the set $S$ is partitioned into several subsets, the join result of the first reducer is incomplete, and another MapReduce is required to combine the results of $R_i \ltimes S_j$ for all $1 \leq j \leq \sqrt{N}$. Therefore, the shuffling cost of H-BRJ is $\sqrt{N} \cdot (|R| + |S|) + \sum_i \sum_j |R_i \ltimes S_j|^2$, while for the basic strategy, it is $|R| + N \cdot |S|$.

In order to reduce the shuffling cost, a better strategy is that $R$ is partitioned into $N$ disjoint subsets and for each subset $R_i$, find a subset of $S_i$ that $R_i \ltimes S = R_i \ltimes S_i$ and $R \ltimes S = \bigcup_{1 \leq i \leq N} R_i \ltimes S_i$. Then, instead of sending the entire $S$ to each reducer (as in the basic strategy) or sending each $R_i$ to $\sqrt{N}$ reducers, $S_i$ is sent to the reducer that $R_i$ belongs to and the $k$NN join is performed between $R_i$ and $S_i$ only.

---

[2] $\sqrt{N} \cdot (|R| + |S|)$ is the shuffling cost of the first MapReduce. $\sum_i \sum_j |R_i \ltimes S_j|$ is the shuffling cost of the second MapReduce for merging the partial results.

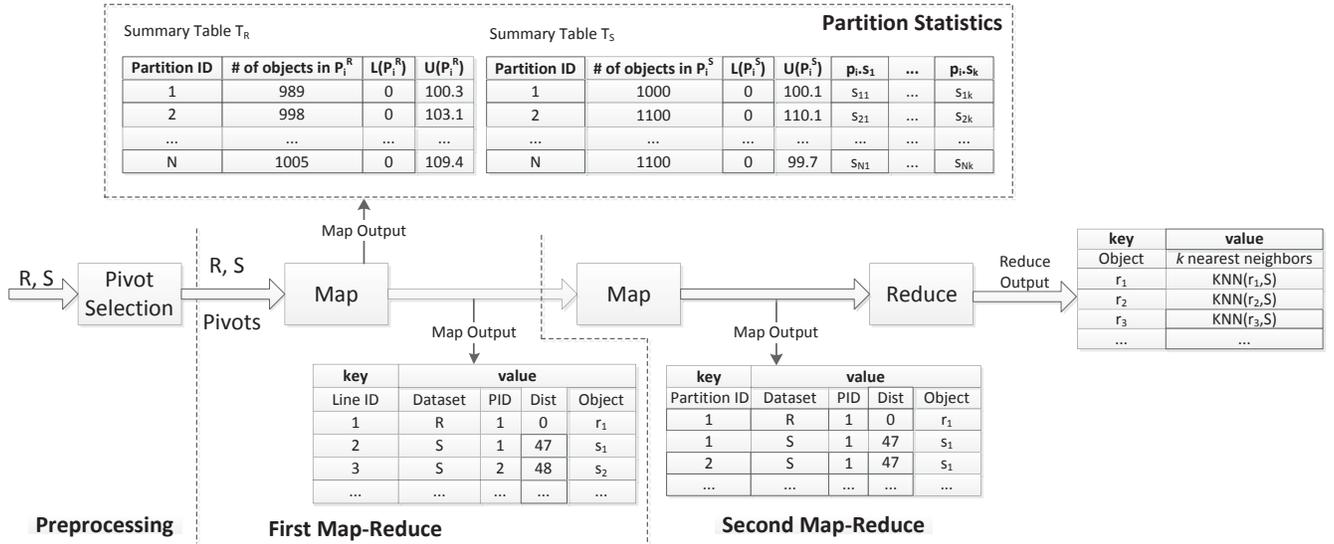

**Figure 3: An overview of $k$NN join in MapReduce**

This approach avoids replication on the set $R$ and sending the entire set $S$ to all reducers. However, to guarantee the correctness of the $k$NN join, the subset $S_i$ must contain the $k$ nearest neighbors of every $r \in R_i$, i.e., $\forall r \in R_i, KNN(r, S) \subseteq S_i$. Note that $S_i \cap S_j$ may not be empty, as it is possible that object $s$ is one of the $k$ nearest neighbors of $r_i \in R_i$ and $r_j \in R_j$. Hence, some of the objects in $S$ should be replicated and distributed to multiple reducers. The shuffling cost is $|R| + \alpha \cdot |S|$, where $\alpha$ is the average number of replicas of an object in $S$. Apparently, if we can reduce the value of $\alpha$, both shuffling and computational cost we consider can be reduced.

In summary, for the purpose of minimizing the join cost, we need to

1. find a good partitioning of $R$;

2. find the minimal set of $S_i$ for each $R_i \in R$, given a partitioning of $R$ [3].

Intuitively, a good partitioning of $R$ should cluster objects in $R$ based on their proximity, so that the objects in a subset $R_i$ are more likely to share common $k$ nearest neighbors from $S$. For each $R_i$, the objects in each corresponding $S_i$ are cohesive, leading to a smaller size of $S_i$. Therefore, such partitioning not only leads to a lower shuffling cost, but also reduces the computational cost of performing the $k$NN join between each $R_i$ and $S_i$, i.e., the number of distance calculations.

## 4. HANDLING KNN JOIN USING MAPREDUCE

In this section, we introduce our implementation of $k$NN join using MapReduce. First, Figure 3 illustrates the working flow of our $k$NN join, which consists of one preprocessing step and two MapReduce jobs.

---

[3] The minimum set of $S_i$ is $S_i = \bigcup_{1 \leq j \leq |R_i|} KNN(r_i, S)$. However, it is impossible to find out the $k$ nearest neighbors for all $r_i$ apriori.

- First, the preprocessing step finds out a set of pivot objects based on the input dataset $R$. The pivots are used to create a Voronoi diagram, which can help partition objects in $R$ effectively while preserving their proximity.

- The first MapReduce job consists of a single Map phase, which takes the selected pivots and datasets $R$ and $S$ as the input. It finds out the nearest pivot for each object in $R \cup S$ and computes the distance between the object and the pivot. The result of the mapping phase is a partitioning on $R$, based on the Voronoi diagram of the pivots. Meanwhile, the mappers also collect some statistics about each partition $R_i$.

- Given the partitioning on $R$, mappers of the second MapReduce job find the subset $S_i$ of $S$ for each subset $R_i$ based on the statistics collected in the first MapReduce job. Finally, each reducer performs the $k$NN join between a pair of $R_i$ and $S_i$ received from the mappers.

### 4.1 Data Preprocessing

As mentioned in previous section, a good partitioning of $R$ for optimizing $k$NN join should cluster objects based on their proximity. We adopt the Voronoi diagram-based data partitioning technique as reviewed in Section 2, which is well-known for maintaining data proximity, especially for data in multi-dimensional space. Therefore, before launching the MapReduce jobs, a preprocessing step is invoked in a master node for selecting a set of pivots to be used for Voronoi diagram-based partitioning. In particular, the following three strategies can be employed to select pivots.

- **Random Selection.** First, $T$ random sets of objects are selected from $R$. Then, for each set, we compute the total sum of the distances between every two objects. Finally, the objects from the set with the maximum total sum distance are selected as the pivots for data partitioning.

- **Farthest Selection.** The set of pivots are selected iteratively from a sample of the original dataset $R$ (since preprocessing procedure is executed on a master node, the original dataset may be too large for it to process). First, we randomly select an object as the first pivot. Next, the object with the largest distance to the first pivot is selected as the second pivot. In the $i^{th}$ iteration, the object that maximizes the sum of its distance to the first $i − 1$ pivots is chosen as the $i^{th}$ pivot.

- **$k$-means Selection.** Similar to the farthest selection, $k$-means selection first does sampling on the $R$. Then, traditional $k$-means clustering method is applied on the sample. With the $k$ data clusters generated, the center point of each cluster is chosen as a pivot for the Voronoi diagram-based data partitioning.

## 4.2 First MapReduce Job

Given the set of pivots selected in the preprocessing step, we launch a MapReduce job for performing data partitioning and collecting some statistics for each partition. Figure 4 shows an example of the input/output of the mapper function of the first MapReduce job.

Specifically, before launching the map function, the selected pivots $\mathbb{P}$ are loaded into main memory in each mapper. A mapper sequentially reads each object $o$ from the input split, computes the distance between $o$ and all pivots in $\mathbb{P}$, and assigns $o$ to the closest pivot $P$. Finally, as illustrated in Figure 4, the mapper outputs each object $o$ along with its partition id, original dataset name ($R$ or $S$), distance to the closest pivot.

Meanwhile, the first map function also collects some statistic for each input data split and these statistics are merged together while the MapReduce job completes. Two in-memory tables called summery tables are created to keep these statistics. Figure 3 shows an example of the summary tables $T_R$ and $T_S$ for partitions of $R$ and $S$, respectively. Specifically, $T_R$ maintains the following information for every partition of $R$: the partition id, the number of objects in the partition, the minimum distance $L(P_i^R)$ and maximum distance $L(P_i^R)$ from an object in partition $P_i^R$ to the pivot. Note that although the pivots are selected based on dataset $R$ alone, the Voronoi diagram based on the pivots can be used to partition $S$ as well. $T_S$ maintains the same fields as those in $T_R$ for $S$. Moreover, $T_S$ also maintains the distances between objects in $KNN(p_i, P_i^S)$ and $p_i$, where $KNN(p_i, P_i^S)$ refers to the $k$ nearest neighbors of pivot $p_i$ from objects in partition $P_i^S$. In Figure 3, $p_i.d_j$ in $T_S$ represents the distance between pivot $p_i$ and its $j^{th}$ nearest neighbor in $KNN(p_i, P_i^S)$. The information in $T_R$ and $T_S$ will be used to guide how to generate $S_i$ for $R_i$ as well as to speed up the computation of $R_i \ltimes S_i$ by deriving distance bounds of the $k$NN for any object of $R$ in the second MapReduce job.

## 4.3 Second MapReduce Job

The second MapReduce job performs the $k$NN join in the way introduced in Section 3. The main task of the mapper in the second MapReduce is to find the corresponding subset $S_i$ for $R_i$. Each reducer performs the $k$NN join between a pair of $R_i$ and $S_i$.

As mentioned previously, to guarantee the correctness, $S_i$ should contain the $k$NN of all $r \in R_i$, i.e., $S_i = \bigcup_{\forall r_j \in R_i} KNN(r_j, S)$. However, we cannot get the exact $S_i$ without performing the $k$NN join on $R_i$ and $S$. Therefore, in the following, we derive a distance bound based on the partitioning of $R$ which can help us reduce the size of $S_i$.

### 4.3.1 Distance Bound of $k$NN

Instead of computing the $k$NN from $S$ for each object of $R$, we derive a bound of the $k$NN distance using a set oriented approach. Given a partition $P_i^R$ (i.e., $R_i$) of $R$, we bound the distance of the $k$NN for all objects of $P_i^R$ at a time based on $T_R$ and $T_S$, which we have as a byproduct of the first MapReduce.

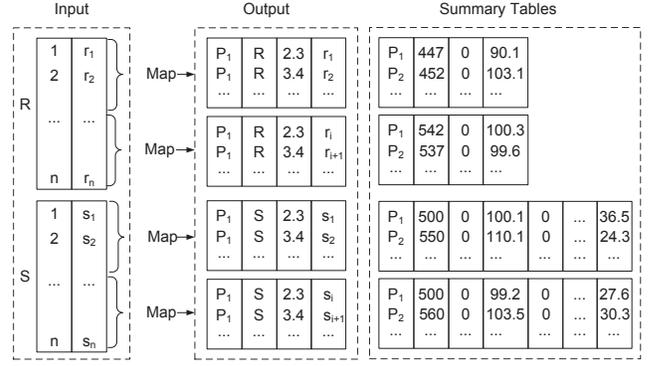

**Figure 4: Partitioning and building the summary tables**

THEOREM 3. *Given a partition $P_i^R \subset R$, an object $s$ of $P_j^S \subset S$, the upper bound distance from $s$ to $\forall r \in P_i^R$, denoted as $ub(s, P_i^R)$, is:*

$$ub(s, P_i^R) = U(P_i^R) + |p_i, p_j| + |p_j, s| \qquad (5)$$

**Proof.** $\forall r \in P_i^R$, according to the triangle inequality, $|r, p_j| \leq |r, p_i| + |p_i, p_j|$. Similarly, $|r, s| \leq |r, p_j| + |p_j, s|$. Hence, $|r, s| \leq |r, p_i| + |p_i, p_j| + |p_j, s|$. Since $r \in P_i^R$, according to the definition of $U(P_i^R)$, $|r, p_i| \leq U(P_i^R)$. Clearly, we can derive $|r, s| \leq U(P_i^R) + |p_i, p_j| + |p_j, s| = ub(s, P_i^R)$. □

Figure 5(a) shows the geometric meaning of $ub(s, P_i^R)$. According to the Equation 5, we can find a set of $k$ objects from $S$ with the smallest upper bound distances as the $k$NN of all objects in $P_i^R$. For ease of exposition, let $KNN(P_i^R, S)$ be the $k$ objects from $S$ with the smallest $ub(s, P_i^R)$. Apparently, we can derive a bound (denoted as $\theta_i$ that corresponds to $P_i^R$) of the $k$NN distance for all objects in $P_i^R$ as follows:

$$\theta_i = \max_{\forall s \in KNN(P_i^R, S)} |ub(s, P_i^R)|. \qquad (6)$$

Clearly, $\forall r \in P_i^R$, the distance from $r$ to any object of $KNN(r, S)$ is less than or equal to $\theta_i$. Hence, we are able to bound the distance of the $k$NN for all objects of $P_i^R$ at a time. Moreover, according to the Equation 5, we can also observe that in each partition $P_j^S$, $k$ objects with the smallest distances to $p_i$ may contribute to refine $KNN(P_i^R, S)$ while the remainder cannot. Hence, we only maintain $k$ smallest distances of objects from each partition of $S$ to its corresponding pivot in summary table $T_S$ (shown in Figure 3).

---

**Algorithm 1:** boundingKNN($P_i^R$)

1   create a priority queue $PQ$;
2   **foreach** $P_j^S$ **do**
3     **foreach** $s \in KNN(p_j, P_j^S)$ **do**   /* set in $T_S$ */
4       $ub(s, P_i^R) \leftarrow U(P_i^R) + |p_i, p_j| + |p_j, s|$;
5       **if** $PQ.size < k$ **then**   $PQ.add(ub(s, P_i^R))$;
6       **else if** $PQ.top > dist$ **then**
7         $PQ.remove()$; $PQ.add(ub(s, P_i^R))$;
8       **else** break;
9   **return** $PQ.top$;

---

Algorithm 1 shows the details on how to compute $\theta_i$. We first create a priority queue $PQ$ with size $k$ (line 1). For partition $P_j^S$, we compute $ub(s, P_i^R)$ for each $s \in KNN(p_j, P_j^S)$, where $|s, p_j|$ is maintained in $T_S$. To speed up the computation of $\theta_i$, we maintain $|s, p_j|$ in $T_S$ based on the ascending order. Hence, when $ub(s, P_i^R) \geq PQ.top$, we can guarantee that no remaining objects in $KNN(p_j, P_j^S)$ help refine $\theta_i$ (line 8). Finally, we return the top of $PQ$ which is taken as $\theta_i$ (line 9).

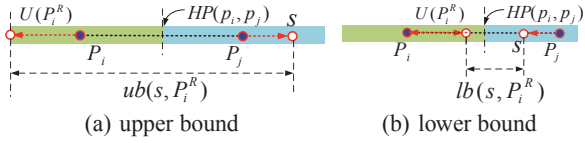

(a) upper bound      (b) lower bound

**Figure 5: Bounding $k$ nearest neighbors**

### 4.3.2 Finding $S_i$ for $R_i$

Similarly to Theorem 3, we can derive the lower bound distance from an object $s \in P_j^S$ to any object of $P_i^R$ as follows.

THEOREM 4. *Given a partition $P_i^R$, an object $s$ of $P_j^S$, the lower bound distance from $s$ to $\forall r \in P_i^R$, denoted by $lb(s, P_i^R)$, is:*

$$lb(s, P_i^R) = \max\{0, |p_i, p_j| - U(P_i^R) - |s, p_j|\} \quad (7)$$

PROOF. $\forall r \in P_i^R$, according to the triangle inequality, $|r, p_j| \geq |p_j, p_i| - |p_i, r|$. Similarly, $|r, s| \geq |r, p_j| - |p_j, s|$. Hence, $|r, s| \geq |p_j, p_i| - |p_i, r| - |p_j, s|$. Since $r \in P_i^R$, according to the definition of $U(P_i^R)$, $|r, p_i| \leq U(P_i^R)$. Thus we can derive $|r, s| \geq |p_i, p_j| - U(P_i^R) - |s, p_j|$. As the distance between any two objects is not less than 0, the low bound distance $lb(s, P_i^R)$ is set to $\max\{0, |p_i, p_j| - U(P_i^R) - |s, p_j|\}$ □

Figure 5(b) shows the geometric meaning of $lb(s, P_i^R)$. Clearly, $\forall s \in S$, if we can verify $lb(s, P_i^R) > \theta_i$, then $s$ cannot be one of $KNN(r, S)$ for any $r \in P_i^R$ and $s$ is safe to be pruned. Hence, it is easy for us to verify whether an object $s \in S$ needs to be assigned to $S_i$.

THEOREM 5. *Given a partition $P_i^R$ and an object $s \in S$, the necessary condition that $s$ is assigned to $S_i$ is that: $lb(s, P_i^R) \leq \theta_i$.*

According to Theorem 5, $\forall s \in S$, by computing $lb(s, P_i^R)$ for all $P_i^R \subset R$, we can derive all $S_i$ that $s$ is assigned to. However, when the number of partitions for $R$ is large, this computation cost might increase significantly since $\forall s \in P_j^S$, we need to compute $|p_i, p_j|$. To cope with this problem, we propose Corollary 2 to find all $S_i$ which $s$ is assigned to only based on $|s, p_j|$.

COROLLARY 2. *Given a partition $P_i^R$ and a partition $P_j^S$, $\forall s \in P_j^S$, the necessary condition that $s$ is assigned to $S_i$ is that:*

$$|s, p_j| \geq LB(P_j^S, P_i^R), \quad (8)$$

*where $LB(P_j^S, P_i^R) = |p_i, p_j| - U(P_i^R) - \theta_i$.*

PROOF. The conclusion directly follows Theorem 5 and Equation 7. □

According to Corollary 2, for partition $P_j^S$, objects exactly lying in region $[LB(P_j^S, P_i^R), U(P_j^S)]$ are assigned to $S_i$. Algorithm 2 shows how to compute $LB(P_j^S, P_i^R)$, which is self-explained.

### 4.3.3 kNN Join between $R_i$ and $S_i$

As a summary, Algorithm 3 describes the details of $k$NN join procedure that is described in the second MapReduce job. Before launching map function, we first compute $LB(P_j^S, P_i^R)$ for every

---

$P_j^S$ (line 1–2). For each object $r \in R$, the map function generates a new key value pair in which the key is its partition id, and the value consists of $k1$ and $v1$ (line 4–6). For each object $s \in S$, the map function creates a set of new key value pairs, if not pruned based on Corollary 2 (line 7–11).

In this way, objects in each partition of $R$ and their potential $k$ nearest neighbors will be sent to the same reducer. By parsing the key value pair ($k2, v2$), the reducer can derive the partition $P_i^R$ and subset $S_i$ that consists of $P_{j_1}^S, \ldots, P_{j_M}^S$ (line 13), and compute the $k$NN of objects in partition $P_i^R$ (line 16–25).

$\forall r \in P_i^R$, in order to reduce the number of distance computations, we first sort the partitions from $S_i$ by the distances from their pivots to pivot $p_i$ in the ascending order (line 14). This is based on the fact that if a pivot is near to $p_i$, then its corresponding partition often has higher probability of containing objects that are closer to $r$. In this way, we can derive a tighter bound distance of $k$NN for every object of $P_i^R$, leading to a higher pruning power. Based on Equation 6, we can derive a bound of the $kNN$ distance, $\theta_i$, for all objects of $P_i^R$. Hence, we can issue a range search with query $r$ and threshold $\theta_i$ over dataset $S_i$. First, $KNN(r, S)$ is set to empty (line 17). Then, all partitions $P_j^S$ are checked one by one (line 18–24). For each partition $P_j^S$, based on Corollary 1, if $d(r, HP(p_i, p_j)) > \theta$, no objects in $P_j^S$ can help refine $KNN(r, S)$, and we proceed to check the next partition directly (line 19–20). Otherwise, $\forall s \in P_j^S$, if $s$ cannot be pruned by Theorem 2, we need to compute the distance $|r, s|$. If $|r, s| < \theta$, $KNN(r, S)$ is updated with $s$ and $\theta$ is updated accordingly (lines 22–24). After checking all partitions of $S_i$, the reducer outputs $KNN(r, S)$ (line 25).

## 5. MINIMIZING REPLICATION OF S

By bounding the $k$ nearest neighbors for all objects in partition $P_i^R$, according to Corollary 2, $\forall s \in P_i^S$, we assign $s$ to $S_i$ when $|s, p_j| \geq LB(P_i^S, P_i^R)$. Apparently, to minimize the number of replicas of objects in $S$, we expect to find a large $LB(P_j^S, P_i^R)$ while keeping a small $|s, p_j|$. Intuitively, by selecting a larger number of pivots, we can split the dataset into a set of Voronoi cells (corresponding to partitions) with finer granularity and the bound the $kNN$ distance for all objects in each partition of $R$ will become tighter. This observation is able to be confirmed by Equation 8. By enlarging the number of pivots, each object from $R \cup S$ is able to be assigned to a pivot with a smaller distance, which reduces both $|s, p_j|$ and the upper bound $U(P_i^R)$ for each partition $P_i^R$ while a smaller $U(P_i^R)$ can help achieve a larger $LB(P_j^S, P_i^R)$. Hence, in order to minimize the replicas of objects in $S$, it is required to select a larger number of pivots. However, in this way, it might not be practical to provide a single reducer to handle each partition $P_i^R$. To cope with this problem, a natural idea is to divide partitions of $R$ into disjoint groups, and take each group as $R_i$. In this way, $S_i$ needs to be refined accordingly.

### 5.1 Cost Model

By default, let $R = \bigcup_{1 \leq i \leq N} G_i$, where $G_i$ is a group consisting of a set of partitions of $R$ and $G_i \cap G_j = \emptyset, i \neq j$.

**Theorem 6.** *Given partition $P_j^S$ and group $G_i$, $\forall s \in P_j^S$, the necessary condition that $s$ is assigned to $S_i$ is:*

$$|s, p_j| \geq LB(P_j^S, G_i), \quad (9)$$

*where $LB(P_j^S, G_i) = \min_{\forall P_i^R \in G_i} LB(P_j^S, P_i^R)$.*

**Proof.** According to Corollary 2, $s$ is assigned to $S_i$ as long as there exists a partition $P_i^R \in G_i$ with $|s, p_j| \geq LB(P_j^S, P_i^R)$. $\square$

By computing $LB(P_j^S, G_i)$ in advance for each partition $P_j^S$, we can derive all $S_i$ for each $s \in P_j^S$ only based on $|s, p_j|$. Apparently, the average number of replicas of objects in $S$ is reduced since duplicates in $S_i$ are eliminated. According to Theorem 6, we can easily derive the number of all replicas (denoted as $RP(S)$) as follows.

**Theorem 7.** *The number of replicas of objects in $S$ that are distributed to reducers is:*

$$RP(S) = \sum_{\forall G_i} \sum_{\forall P_j^S} |\{s | s \in P_j^S \land |s, p_j| \geq LB(P_j^S, G_i)\}| \quad (10)$$

### 5.2 Grouping Strategies

We present two strategies for grouping partitions of $R$ to approximately minimize $RP(S)$.

<hr/>

**Algorithm 4:** geoGrouping()

1 select $p_i$ such that $\sum_{\forall p_j \in \mathbb{P}} |p_i, p_j|$ is maximized;
2 $\tau \leftarrow \{p_i\}$; $G_1 \leftarrow \{P_i^R\}$; $\mathbb{P} \leftarrow \mathbb{P} - \{p_i\}$;
3 **for** $i \leftarrow 2$ **to** $N$ **do**
4    select $p_l \in \mathbb{P}$ such that $\sum_{\forall p_j \in \tau} |p_j, p_l|$ is maximized;
5    $G_i \leftarrow \{P_l^R\}$; $\mathbb{P} \leftarrow \mathbb{P} - \{p_l\}$; $\tau \leftarrow \tau \cup \{p_l\}$;
6 **while** $\mathbb{P} \neq \emptyset$ **do**
7    select group $G_i$ with the smallest number of objects;
8    select $p_l \in \mathbb{P}$ such that $\sum_{\forall P_j^R \in G_i} |p_l, p_j|$ is minimized;
9    $G_i \leftarrow G_i \cup \{P_l^R\}$; $\mathbb{P} \leftarrow \mathbb{P} - \{p_l\}$;
10 **return** $\{G_1, G_2, \ldots, G_N\}$

<hr/>

#### 5.2.1 Geometric Grouping

Geometric grouping is based on an important observation: given two partitions $P_i^R$ and $P_j^S$, if $p_j$ is far away from $p_i$ compared with the remaining pivots, then $P_j^S$ is deemed to have a low possibility of containing objects as any of $kNN$ for objects in $P_i^R$. This observation can be confirmed in Figure 1 where partition $P_5$ does not have objects to be taken as any of $kNN$ in $P_2$. Hence, a natural idea to divide partitions of $R$ is that we make the partitions, whose corresponding pivots are near to each other, into the same group. In this way, regarding group $G_i$, objects of partitions from $S$ that are far away from partitions of $G_i$ will have a large possibility to be pruned.

Algorithm 4 shows the details of geometric grouping. We first select the pivot $p_i$ with the farthest distance to all the other pivots (line 1) and assign partition $P_i^R$ to group $G_1$ (line 2). We then sequentially assign a partition to the remaining groups as follows: for group $G_i$ ($2 \leq i \leq N$), we compute the pivot $p_l$ which has the farthest distance to the selected pivots ($\tau$) and assign $P_l^R$ to $G_i$ (line 3–5). In this way, we can guarantee that the distance among all groups are the farthest at the initial phase. After assigning the first partition for each group, in order to balance the workload, we do the following iteration until all partitions are assigned to the groups: (1) select the group $G_i$ with the smallest number of objects (line 7); (2) compute the pivot $p_l$ with the minimum distance to the pivots of $G_i$, and assign $P_l^R$ to $G_i$ (line 8–9). In this way, we can achieve that the number of objects in each group is nearly the same. Finally, we return all groups that maintain partitions of $R$ (line 10).

#### 5.2.2 Greedy Grouping

Let $RP(S, G_i)$ be the set of objects from $S$ that need to be assigned to $S_i$. The objective of greedy grouping is to minimize the size of $RP(S, G_i \cup \{P_j^R\}) - RP(S, G_i)$ when assigning a new partition $P_j^R$ to $G_i$. According to Theorem 6, $RP(S, G_i)$ is able to be formally quantified as:

$$RP(S, G_i) = \bigcup_{\forall P_j^S \subset S} \{s | s \in P_j^S \land |s, p_j| \geq LB(P_j^S, G_i)\} \quad (11)$$

Hence, theoretically, when implementing the greedy grouping approach, we can achieve the optimization objective by minimizing $RP(S, G_i \cup \{P_j^R\}) - RP(S, G_i)$ instead of $\sum_{\forall P_j^R \in G_i} |p_i, p_j|$ in the geometric grouping approach. However, it is rather costly to select a partition $P_j^R$ from all remaining partitions with minimum $RP(S, G_i \cup \{P_j^R\}) - RP(S, G_i)$. This is because by adding a new partition $P_j^R$ to $G_i$, we need to count the number of emerging objects from $S$ that are distributed to the $S_i$. Hence, to reduce the computation cost, once $\exists s \in P_l^S$, $|s, p_j| \leq LB(P_j^S, G_i)$, we add all objects of partition $P_i^S$ to $RP(S, G_i)$, i.e., the $RP(S, G_i)$ is approximately quantified as:

$$RP(S, G_i) \approx \bigcup_{\forall P_j^S \subset S} \{P_j^S | LB(P_j^S, G_i) \leq U(P_j^S)\} \qquad (12)$$

**Remark:** To answer $k$NN join by exploiting the grouping strategies, we use the group id as the key of the Map output. We omit the details which are basically the same as described in Algorithm 3.

# 6. EXPERIMENTAL EVALUATION

We evaluate the performance of the proposed algorithms on our in-house cluster, Awan[4]. The cluster includes 72 computing nodes, each of which has one Intel X3430 2.4GHz processor, 8GB of memory, two 500GB SATA hard disks and gigabit ethernet. On each node, we install CentOS 5.5 operating system, Java 1.6.0 with a 64-bit server VM, and Hadoop 0.20.2. All the nodes are connected via three high-speed switches. To adapt the Hadoop environment to our application, we make the following changes to the default Hadoop configurations: (1) the replication factor is set to 1; (2) each node is configured to run one map and one reduce task. (3) the size of virtual memory for each map and reduce task is set to 4GB.

We evaluate the following approaches in the experiments.

- H-BRJ is proposed in [21] and described in Section 3. In particular, to speed up the computation of $R_i \bowtie S_j$, it employs R-tree to index objects of $S_j$ and finds $k$NN for $\forall r \in R_i$ by traversing the R-tree. We used the implementation generously provided by the authors;

- PGBJ is our proposed $k$NN join algorithm that utilizes the partitioning and grouping strategy;

- PBJ is also our proposed $k$NN join algorithm. The only difference between PBJ and PGBJ is that PBJ does not have the grouping part. Instead, it employs the same framework used in H-BRJ. Hence, it also requires an extra MapReduce job to merge the final results.

We conduct the experiments using self-join on the following datasets:

- Forest FCoverType[5] (Forest for short): This is a real dataset that predicts forest cover type from cartographic variables. There are 580K objects, each with 54 attributes (10 integer, 44 binary). We use 10 integer attributes in the experiments.

- Expanded Forest FCoverType dataset: To evaluate the performance on large datasets, we increase the size of Forest while maintaining the same distribution of values over the dimensions of objects (like [16]). We generate new objects in the way as follows: (1) we first compute the frequencies of values in each dimension, and sort values in the ascending order of their frequencies; (2) for each object $o$ in the original dataset, we create a new object $\bar{o}$, where in each dimension $D_i$, $\bar{o}[i]$ is ranked next to $o[i]$ in the sorted list. Further, to create multiple new objects based on object $o$, we replace $o[i]$ with a set of values next to it in the sorted list for $D_i$. In particular, if $o[i]$ is the last value in the list for $D_i$, we keep this value constant. We build Expanded Forest FCoverType dataset by increasing the size of Forest dataset from 5 to 25 times. We use "Forest $\times t$" to denote the increased dataset where $t \in [5, 25]$ is the increase factor.

- OpenStreetMap[6] (OSM for short): this is a real map dataset containing the location and description of objects. We extract 10 million records from this dataset, where each record consists of 2 real values (longitude and latitude) and a description with variable length.

By default, we evaluate the performance of $k$NN join ($k$ is set to 10) on the "Forest $\times 10$" dataset using 36 computing nodes. We measure several parameters, including query time, distance computation selectivity, and shuffling cost. The distance computation selectivity (computation selectivity for short) is computed as follows:

$$\frac{\text{\# of object pairs to be computed}}{|R| \times |S|}, \qquad (13)$$

where the objects also include the pivots in our case.

## 6.1 Study of Parameters of Our Techniques

We study the parameters of PGBJ with respect to pivot selection strategy, pivot number, and grouping strategy. By combining different pivot selection and grouping strategies, we obtain 6 strategies, which are: (1) **RGE**, random selection + geometric grouping; (2) **FGE**, farthest selection + geometric grouping; (3) **KGE**, $k$-means selection + geometric grouping; (4) **RGR**, random selection + greedy grouping; (5)**FGR**, farthest selection + greedy grouping; (6) **KGR**, $k$-means selection + greedy grouping.

### 6.1.1 Effect of Pivot Selection Strategies

Table 2 shows the statistics of partition sizes using different pivot selection strategies including random selection, farthest selection and $k$-means selection. We observe that the standard deviation (`dev.`for short) of partition size drops rapidly when the number of pivots increases. Compared to random selection and $k$-means selection, partition size varies significantly using the farthest selection. The reason is that in the farthest selection, outliers are always selected as pivots. Partitions corresponding to these pivots contain few objects, while other partitions whose pivots reside in dense areas contain a large number of objects. Specifically, when we select 2000 pivots using farthest selection, the maximal partition size is 1,130,678, which is about 1/5 of the dataset size. This large difference in partition size will degrade performance due to the unbalanced workload. We also investigate the group size using geometric grouping approach[7]. As shown in Table 3, the number of objects in each group varies significantly using the farthest selection. Again, this destroys the load balance since each reducer needs to perform significantly different volume of computations. However, the group sizes using random selection and $k$-means selection are approximately the same.

Figure 6 shows the execution time for various phases in $k$NN join. We do not provide the execution time for farthest selection because it takes more than 10,000s to answer $k$NN join. The reason of the poor performance is: almost all the partitions of $S$ overlap with large-size partitions of $R$. Namely, we need to compute distances for a large number of object pairs. Comparing **RGE** with **KGE**, and **RGR** with **KGR** in Figure 6, we observe that the overall performance using random selection is better than that using $k$-means selection. Further, when the number of pivots increases, the gap of the overall performance becomes larger. This is because $k$-means selection involves a large number of distance computations, which results in large execution time. Things get worse when $k$ increases.

**Table 2: Statistics of partition size**

| # of pivots | Random Selection | | | | Farthest Selection | | | | k-means Selection | | | |
|---|---|---|---|---|---|---|---|---|---|---|---|---|
| | min. | max. | avg. | dev. | min. | max. | avg. | dev. | min. | max. | avg. | dev. |
| 2000 | 116 | 9062 | 2905.06 | 1366.50 | 24 | 1130678 | 2905.06 | 27721.10 | 52 | 7829 | 2905.06 | 1212.38 |
| 4000 | 116 | 5383 | 1452.53 | 686.41 | 14 | 1018605 | 1452.53 | 13313.56 | 17 | 5222 | 1452.53 | 700.20 |
| 6000 | 24 | 4566 | 968.35 | 452.79 | 13 | 219761 | 968.35 | 5821.18 | 3 | 3597 | 968.35 | 529.92 |
| 8000 | 6 | 2892 | 726.27 | 338.88 | 12 | 97512 | 726.27 | 2777.84 | 6 | 2892 | 726.27 | 338.88 |

**Table 3: Statistics of group size**

| # of pivots | Random Selection | | | | Farthest Selection | | | | k-Means Selection | | | |
|---|---|---|---|---|---|---|---|---|---|---|---|---|
| | min. | max. | avg. | dev. | min. | max. | avg. | dev. | min. | max. | avg. | dev. |
| 2000 | 143720 | 150531 | 145253 | 1656 | 86805 | 1158084 | 145253 | 170752 | 143626 | 148111 | 145253 | 1201 |
| 4000 | 144564 | 147180 | 145253 | 560 | 126635 | 221539 | 145253 | 20204 | 144456 | 146521 | 145253 | 570 |
| 6000 | 144758 | 146617 | 145253 | 378 | 116656 | 1078712 | 145253 | 149673 | 144746 | 155858 | 145253 | 342 |
| 8000 | 144961 | 146118 | 145253 | 251 | 141072 | 173002 | 145253 | 6916 | 144961 | 146118 | 145253 | 251 |

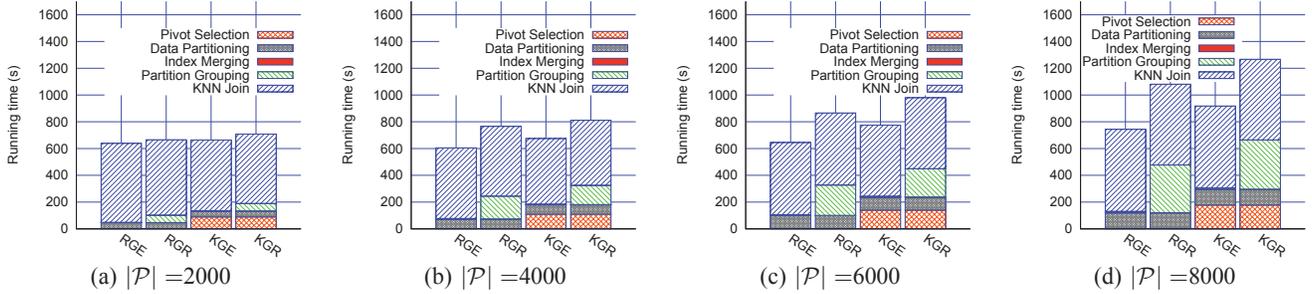

(a) $|\mathcal{P}|$ =2000    (b) $|\mathcal{P}|$ =4000    (c) $|\mathcal{P}|$ =6000    (d) $|\mathcal{P}|$ =8000

**Figure 6: Query cost of tuning parameters**

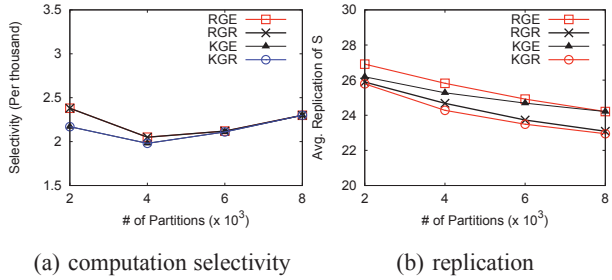

(a) computation selectivity    (b) replication

**Figure 7: Computation selectivity & replication**

However, during the $k$NN join phase, the performance of $k$-means selection is slightly better than that of random selection. To verify the result, we investigate the computation selectivity for both cases. As shown in Figure 7(a), we observe that the computation selectivity of using $k$-means selection is less than that of using random selection. Intuitively, $k$-means selection is more likely to select high-quality pivots that separate the whole dataset more evenly, which enhances the power of our pruning rules. However, another observation is that the selectivity difference becomes smaller when the number of pivots increases. This is because $k$-means selection will deteriorate into random selection when the number of pivots becomes larger. It is worth mentioning that the computation selectivity of all the techniques is low, where the maximum is only 2.38‰.

### 6.1.2 Effect of the Pivot Number

From Figure 6, we observe that the minimal execution time for

$k$NN join phase occurs when $|\mathcal{P}| = 4000$. To specify the reason, we provide the computation selectivity in Figure 7(a). From this figure, we find that the computation selectivity drops by varying $|\mathcal{P}|$ from 2000 to 4000, but increases by varying $|\mathcal{P}|$ from 4000 to 8000. As discussed in $k$NN join algorithm, to compute $KNN(r, S)$, we need to compute the distances between $r$ and objects from $S$ as well as between $r$ and $p_i \in \mathcal{P}$. When the number of pivots increases, the whole space will be split into a finer granularity and the pruning power will be enhanced as the bound becomes tighter. This leads to a reduction in both distance computation between $R$ and $S$ and replication for $S$. The results for replication of $S$ are shown in Figure 7(b). One drawback of using a large number of pivots is that the number of distance computation between $r$ and the pivots becomes larger. On balance, the computation selectivity is minimized when $|\mathcal{P}| = 4000$. For the overall execution time, it arrives at the minimum value when $|\mathcal{P}| = 4000$ for **RGE** and $|\mathcal{P}| = 2000$ for the remaining strategies. The overall performance degrades for all the combination of pivot selection and partition grouping strategies when the number of pivots increases.

### 6.1.3 Effect of Grouping Strategies

When comparing **RGE** with **RGR**, and **KGE** with **KGR** in Figure 6, we find the execution time in the $k$NN join phase remains almost the same using different grouping strategies. In fact, in our partitioning based approach, for each object $r$ with all its potential $k$ nearest neighbors, the number of distance computations for $r$ remains constant. This is consistent with the results for the number of object pairs to be computed in Figure 7(a). As described above, in PGBJ, $\forall r \in R_i$, we send all its potential $k$NN from $S$ to the same reducer. Hence, the shuffling cost depends on how to partition $R$ into subsets. From Figure 7(b), when $|\mathcal{P}|$ increases, the

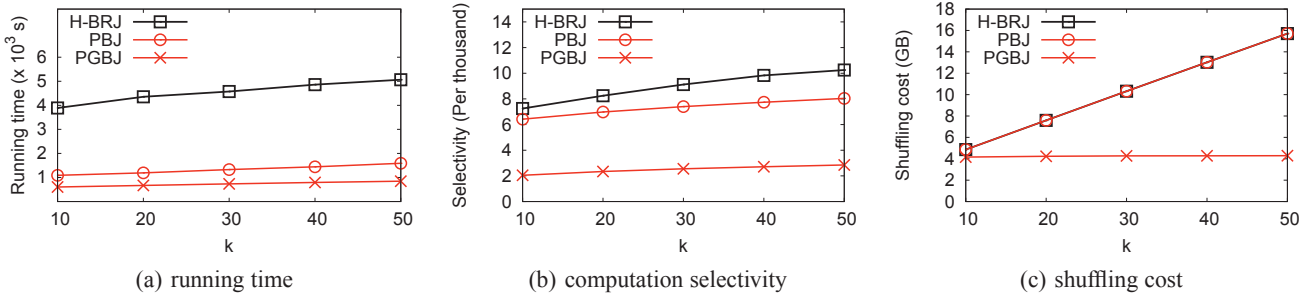

**(a) running time**  **(b) computation selectivity**  **(c) shuffling cost**

**Figure 8: Effect of $k$ over "Forest $\times$ 10"**

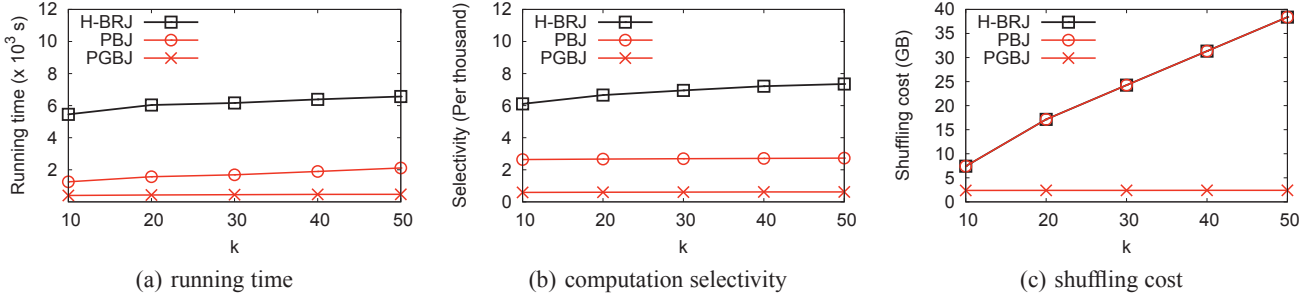

**(a) running time**  **(b) computation selectivity**  **(c) shuffling cost**

**Figure 9: Effect of $k$ over OSM dataset**

average replication of $S$ using greedy grouping is slightly reduced. However, the execution time in partition grouping phase increases significantly. This leads to the increment in the overall execution time.

**Remark.** To summarize the study of the parameters, we find that the overall execution time is minimized when $|\mathcal{P}| = 4000$ and **RGE** strategy is adopted to answer $k$NN join. Hence, in the remaining experiments, for both PBJ and PGBJ, we randomly select 4000 pivots to partition the datasets. Further, we use geometric grouping strategy to group the partitions for PBGJ.

### 6.2 Effect of $k$

We now study the effect of $k$ on the performance of our proposed techniques. Figure 8 and Figure 9 present the results by varying $k$ from 10 to 50 on "Forest $\times$ 10" and OSM datasets, respectively.

In terms of running time, PGBJ always performs best, followed by PBJ and H-BRJ. This is consistent with the results for computation selectivity. H-BRJ requires each reducer to build a R-tree index for all the received objects from $S$. To find the $k$NN for an object from $R$, the reducers will traverse the index and maintain candidate objects as well as a set of intermediate nodes in a priority queue. Both operations are costly for multi-dimensional objects, which result in the long running time. In PGJ, our proposed pruning rules allow each reducer to derive a distance bound from received objects in $S$. This bound is used to reduce computation cost for $k$NN join. However, without grouping phase, PGJ randomly sends a subset of $S$ to each reducer. This randomness results in a loose distance bound, thus degrading the performance. In addition, Figure 8(c) shows the shuffling cost of three approaches on the default dataset. As we can see, when $k$ increases, the shuffling cost of PGBJ remains nearly the same, while it increases linearly for PBJ and H-BRJ. This indicates that the replication of $S$ in PGBJ is insensitive to $k$. However, for H-BRJ and PBJ, the shuffling cost of $R_i \bowtie S_j$ ($\forall R_i \subset R, S_j \subset S$) increases linearly when $k$ varies.

### 6.3 Effect of Dimensionality

We now evaluate the effect of dimensionality. Figure 10 presents both the running time and computation selectivity by varying the number of dimensions from 2 to 10.

From the results, we observe that H-BRJ is more sensitive to the number of the dimensions than PBJ and PGBJ. In particular, the execution time increases exponentially when $n$ varies from 2 to 6. This results from the curse of dimensionality. When the number of dimensions increases, the number of object pairs to be computed increases exponentially. Interestingly, the execution time of $k$NN join increases smoothly when $n$ varies from 6 to 10. To explain this phenomenon, we analyze the original dataset and find that values of 6–10 attributes have low variance, which means the $k$NN for objects from $R$ do not change too much by adding these dimensions. We show the shuffling cost in Figure 10(c). For H-BRJ and PBJ, when the number of dimensions increases, the shuffling cost increases linearly due to the larger data size. However, for PGB-J, when the number of dimensions varies from 2 to 6, the shuffling cost increases exponentially due to the exponential increment of the replication of $S$. Nevertheless, it will converge to $|R| + N \times |S|$ even at the worst case. Although it may exceed both H-BRJ and PBJ, in that case, PBJ can be used instead of PBGJ if we take the shuffling cost into main consideration.

### 6.4 Scalability

We now investigate the scalability of three approaches. Figure 11 presents the results by varying the data size from 1 to 25 times of the original dataset.

From Figure 11(a), we can see that the overall execution time of all the three approaches quadratically increases when we enlarge the data size. This is determined by the fact that the number of object pairs increase quadratically with the data size. However, PGBJ scales better than both PBJ and H-BRJ. In particular, when data size becomes larger, the running time of PGBJ grows much slower than that of H-BRJ. To verify this result, we analyze the computation selectivity for the three approaches. As shown in Figure 11(b), the computation selectivity of PGBJ is always the smallest one. One observation is that when data size increases, the selectivity differences among three approaches tend to be constant. In practice, for large datasets with multi-dimensional objects, a tiny decrease in selectivity will lead to a dramatic improvement in performance. This is the reason that the running time of PGBJ is nearly 6 times faster than that of H-BRJ on "Forest × 25", even if their selectivity does not differ too much. We also present the shuffling cost in Figure 11(c). From the figure, we observe that the shuffling cost of PGBJ is still less than that of PBJ and H-BRJ, and there is an obvious trend of increasing returns when the data size increases.

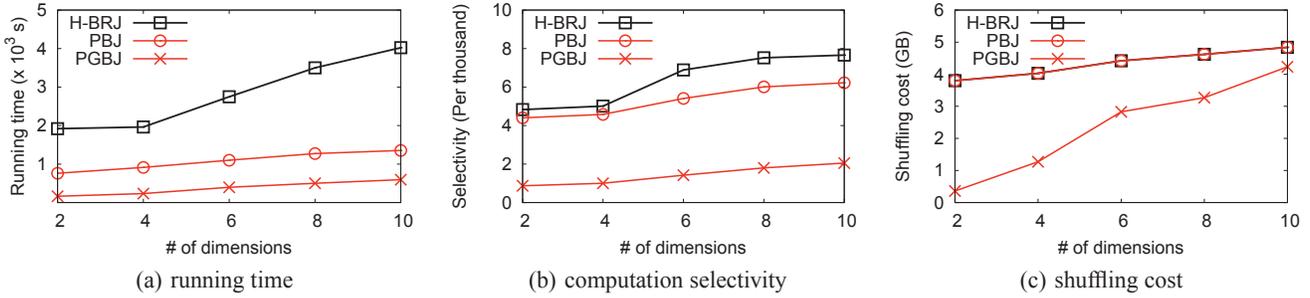

**Figure 10: Effect of dimensionality**

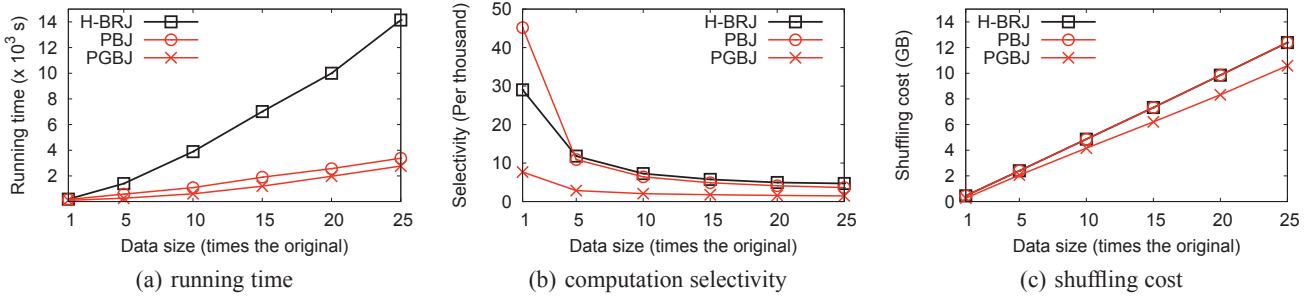

**Figure 11: Scalability**

## 6.5 Speedup

We now measure the effect of the number of computing nodes. Figure 12 presents the results by varying the number of computing nodes from 9 to 36.

From Figure 12(a), we observe that the gap of running time among three approaches tends to be smaller when the number of computing nodes increases. Due to the increment of number of computing nodes, for H-BRJ and PBJ, the distribution of objects over each reducer becomes sparser. This leads to an increment of computation selectivity that is shown in Figure 12(b). However, the computation selectivity for PGBJ remains constant. Based on this trend, it is reasonable to expect that PGBJ will always outperform both H-BRJ and PBJ, while the improvement in running time is getting less obvious. We also show the shuffling cost in Figure 12(c). From the figure, we can see that the shuffling cost increases linearly with the number of computing nodes. In addition, our approaches cannot speed up linearly, because: (1) each node needs to read pivots from the distributed file system; (2) the shuffling cost will be increased.

## 7. RELATED WORK

In centralized systems, various approaches based on the existing indexes have been proposed to answer $k$NN join. In [3, 2], they propose **Mux**, a R-tree based method to answer $k$NN join. It organizes the input datasets with large-sized pages to reduce the I/O cost. Then, by carefully designing a secondary structure with much smaller size within pages, the computation cost is reduced as well. Xia et al. [17] propose a grid partitioning based approach named **Gorder** to answer $k$NN join. Gorder employs the Principal Components Analysis (PCA) technique on the input datasets and sorts the objects according to the proposed Grid Order. Objects are then assigned to different grids where objects in close proximity always lie in the same grid. Finally, it applies the scheduled block nested loop join on the grid data so as to reduce both CPU and I/O costs. Yu et al. [19] propose **IJoin**, a B$^+$-tree based method to answer $k$NN join. Similar to our proposed methods, by splitting the two input datasets into respective set of partitions, IJoin method employs a B$^+$-tree to maintain the objects of each dataset using the iDistance technique [20, 9] and answer $k$NN join based on the properties of B$^+$-tree. Yao et al. [18] propose **Z-KNN**, a Z-ordering based method to answer $k$NN join in relational RDBMS by SQL operators without changes to the database engine. Z-KNN method transforms the $k$NN join operation into a set of $k$NN search operations with each object of $R$ as a query point.

Recently, there has been considerable interest on supporting similarity join queries over MapReduce framework. In [16, 13], they study how to perform set-similarity join in parallel using MapReduce. Set-similarity join returns all object pairs whose similarity does not exceed a given threshold, given the similarity function like Jaccard. Due to the different problem definitions, it is not applicable to extend their techniques to solve our problem. Similar to our methods, Akdogan et al. [1] adopt the Voronoi diagram partitioning based approach using MapReduce to answer range search and $k$NN search queries. In their method, they take each object of the dataset as a pivot and utilize their pivots to partition the space. Obviously, it incurs high maintenance cost and computation cost when the dimension increases. In their work, they claim their method limits to handle 2-dimensional datasets. More related study to our work appears in [14], which proposes a general framework for processing join queries with arbitrary join conditions using MapReduce. Under their framework, they propose various optimization techniques to minimize the communication cost. Although we have different motivations, it is still interesting to extend our methods to their framework in the further work. In [11], they study how to extract $k$ closest object pairs from two input datasets in MapReduce, which is the special case of our proposed problem. In particular, we focus on exactly processing $k$NN join queries in this paper, thus excluding approximate methods, like LSH [7, 15], or H-zkNNJ [21].

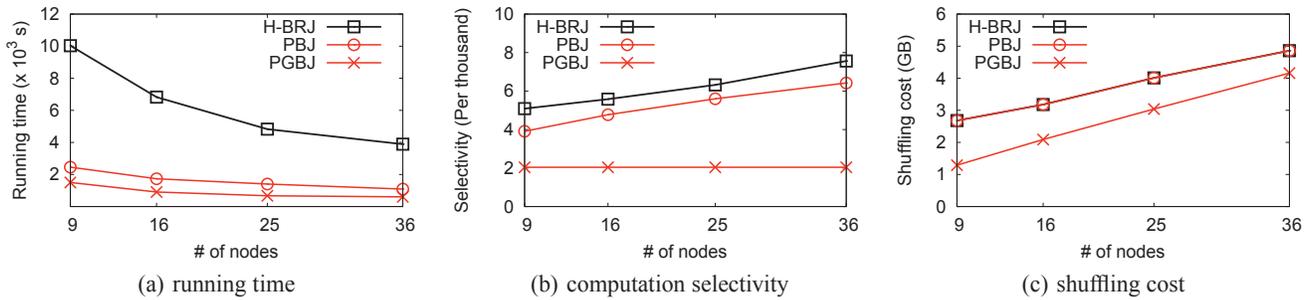

(a) running time      (b) computation selectivity      (c) shuffling cost

**Figure 12: Speedup**

## 8. CONCLUSION

In this paper, we study the problem of efficiently answering the $k$ nearest neighbor join using MapReduce. By exploiting Voronoi diagram-based partitioning method, our proposed approach is able to divide the input datasets into groups and we can answer the $k$ nearest neighbor join by only checking object pairs within each group. Several pruning rules are developed to reduce the shuffling cost as well as the computation cost. Extensive experiments performed on both real and synthetic datasets demonstrate that our proposed methods are efficient, robust and scalable.

### Acknowledgments

The work in this paper was in part supported by the Singapore Ministry of Education Grant No. R-252-000-454-112. We thank Professor Feifei Li for providing us the implementation of their algorithm used in the experiments.

## 9. REFERENCES

[1] A. Akdogan, U. Demiryurek, F. B. Kashani, and C. Shahabi. Voronoi-based geospatial query processing with MapReduce. In *CloudCom*, pages 9–16, 2010.

[2] C. Böhm and F. Krebs. Supporting KDD applications by the k-nearest neighbor join. In *DEXA*, pages 504–516, 2003.

[3] C. Böhm and F. Krebs. The k-nearest neighbour join: Turbo charging the KDD process. *Knowl. Inf. Syst.*, 6(6):728–749, 2004.

[4] C. Böhm and H.-P. Kriegel. A cost model and index architecture for the similarity join. In *ICDE*, pages 411–420, 2001.

[5] M. M. Breunig, H.-P. Kriegel, R. T. Ng, and J. Sander. Lof: Identifying density-based local outliers. In *SIGMOD*, pages 93–104, 2000.

[6] J. Dean and S. Ghemawat. MapReduce: Simplified data processing on large clusters. In *OSDI*, pages 137–150, 2004.

[7] A. Gionis, P. Indyk, and R. Motwani. Similarity search in high dimensions via hashing. In *VLDB*, pages 518–529, 1999.

[8] G. R. Hjaltason and H. Samet. Index-driven similarity search in metric spaces. *ACM Trans. Database Syst.*, 28(4):517–580, 2003.

[9] H. V. Jagadish, B. C. Ooi, K.-L. Tan, C. Yu, and R. Zhang. idistance: An adaptive $B^+$-tree based indexing method for nearest neighbor search. *ACM Trans. Database Syst.*, 30(2):364–397, 2005.

[10] D. Jiang, B. C. Ooi, L. Shi, and S. Wu. The performance of MapReduce: An in-depth study. *PVLDB*, 3(1):472–483, 2010.

[11] Y. Kim and K. Shim. Parallel top-k similarity join algorithms using MapReduce. In *ICDE*, 2012.

[12] E. M. Knorr and R. T. Ng. Algorithms for mining distance-based outliers in large datasets. In *VLDB*, pages 392–403, 1998.

[13] A. Metwally and C. Faloutsos. V-smart-join: A scalable MapReduce framework for all-pair similarity joins of multisets and vectors. *PVLDB*, 5(8):704–715, 2012.

[14] A. Okcan and M. Riedewald. Processing theta-joins using MapReduce. In *SIGMOD*, pages 949–960, 2011.

[15] A. Stupar, S. Michel, and R. Schenkel. RankReduce - processing k-nearest neighbor queries on top of MapReduce. In *LSDS-IR*, pages 13–18, 2010.

[16] R. Vernica, M. J. Carey, and C. Li. Efficient parallel set-similarity joins using MapReduce. In *SIGMOD*, pages 495–506, 2010.

[17] C. Xia, H. Lu, B. C. Ooi, and J. Hu. Gorder: An efficient method for knn join processing. In *VLDB*, pages 756–767, 2004.

[18] B. Yao, F. Li, and P. Kumar. K nearest neighbor queries and knn-joins in large relational databases (almost) for free. In *ICDE*, pages 4–15, 2010.

[19] C. Yu, B. Cui, S. Wang, and J. Su. Efficient index-based knn join processing for high-dimensional data. *Information and Software Technology*, 49(4):332–344, 2007.

[20] C. Yu, B. C. Ooi, K.-L. Tan, and H. V. Jagadish. Indexing the distance: An efficient method to knn processing. In *VLDB*, pages 421–430, 2001.

[21] C. Zhang, F. Li, and J. Jestes. Efficient parallel knn joins for large data in MapReduce. In *EDBT*, 2012.